\documentclass[10pt]{iopart}
\usepackage{graphicx}
\usepackage{xspace}				
\usepackage[pdftex]{color}			
\usepackage{pdfcolmk}				

\newcommand{\bma}[1]{\mbox{\boldmath$#1$}}

\newcommand{\esa}{\textsl{ESA}\xspace}

\newcommand{\lisa}{\textsl{LISA}\xspace}
\newcommand{\lpf}{\textsl{\mbox{LPF}}\xspace}
\newcommand{\ltp}{\textsl{LTP}\xspace}
\newcommand{\nasa}{\textsl{NASA}\xspace}

\begin{document}


\jl{6}

\title[Thermal diagnostic of the Optical Window\ldots]{Thermal diagnostic
of the Optical Window on board \textsl{LISA PathFinder}}

\author{M~Nofrarias$^1$, AF~Garc\'ia Mar\'in$^2$, A~Lobo$^{1,3}$,
G~Heinzel$^2$, J~Ramos-Castro$^4$, J~Sanju\'an$^1$ and K~Danzmann$^2$}

\address{$^1$ Institut d'Estudis Espacials de Catalunya (IEEC),
 Edifici Nexus, Gran Capit\`a 2-4, 08034 Barcelona, Spain}

\address{$^2$Max-Planck-Institut f\"ur Gravitationsphysik
 (Albert-Einstein-Institut), Callinstrasse 38, D-30167 Hannover, Germany}

\address{$^3$ Institut de Ci\`encies de l'Espai, {\sl CSIC}, Facultat de
 Ci\`encies, Torre C5 parell, 08193 Bellaterra, Spain}

\address{$^4$Departament d'Enginyeria Electr\`onica, UPC, Campus Nord,
 Edifici C4, Jordi Girona 1-3, 08034 Barcelona, Spain}
 \ead{nofraria@ieec.uab.es}

\date{11-June-2007}

\begin{abstract}
Vacuum conditions inside the \ltp Gravitational Reference Sensor must be
under 10$^{-5}$ Pa, a rather demanding requirement. The \emph{optical window}
(OW) is an interface which seals the vacuum enclosure and, at the same
time, lets the laser beam go through for interferometric metrology with
the test masses. The OW is a plane-parallel plate clamped in a Titanium
flange, and is considerably sensitive to thermal and stress fluctuations.
It is critical for the required precision measurements, hence its
temperature will be carefully monitored in flight. This paper reports on
the results of a series of OW characterisation laboratory runs, intended
to study its response to selected thermal signals, as well as their fit
to numerical models, and the meaning of the latter. We find that a single
pole \textsl{ARMA} transfer function provides a consistent approximation
to the OW response to thermal excitations, and derive a relationship with
the physical processes taking place in the OW. We also show how system
noise reduction can be accomplished by means of that transfer function.
\end{abstract}
\submitto{\CQG}
%

\section{Introduction}
\label{sec.1}

\lisa Pathfinder (\lpf) is an \esa mission, with \nasa contributions,
whose main objective is to put to test critical parts of \lisa (Laser
Interferometer Space Antenna), the first space borne gravitational wave
(GW) observatory~\cite{bender}. The science module on board \lpf is the
\lisa Technology Package (\ltp)~\cite{lpfall}. The unprecedented
sensitivity of the \ltp has prompted the conceptual enhancement of
\lpf's science objectives as regards the purity of geodesic, or
\emph{free fall} motion of test masses in the interplanetary
gravitational field~\cite{case,nored}.

Free fall control is achieved by Gravitational Reference Sensors
(GRS)~\cite{rita}. These are a set of capacitive sensors which can
determine to high precision (nano-metres) the 3D position and orientation
of cubic test masses relative to their non-contacting enclosure, which is
rigidly linked to the spacecraft structure. Detected off-centre deviations
trigger action by a set of micro-thrusters which \emph{move the spacecraft}
such that the mass returns to its centred position. The combination of the
GRS, the thrusters and the control system (the DFACS, Drag Free and
Attitude Control System) is called \emph{drag-free} subsystem. The latter
is intended to accurately nullify the effects of any non-gravitational forces
acting on the spacecraft. This makes possible to detect \emph{differential}
gravitational accelerations between the two test masses, whether by
precision interferometry~\cite{gerhar} or by the drag free system itself.

This is fundamental for \lisa, since Gravitational Waves (GWs) show
up as \emph{tides}, i.e., time varying differential gravitational
accelerations. The precision of the measurement done with the \ltp is
required to be~\cite{toplev}
\begin{equation}
 \hspace*{-6em}
 S_{\Delta a}^{1/2}(\omega)\leq 3\!\times\!10^{-14}\,\left[
 1 + \left(\frac{\omega/2\pi}{3\ {\rm mHz}}\right)^{\!\!2}\right]\,
 {\rm m}\,{\rm s}^{-2}/\sqrt{\rm Hz}\ ,\quad
 1\,{\rm mHz}\leq\frac{\omega}{2\pi}\leq 30\,{\rm mHz}
 \label{eq.1}
\end{equation}

We shall refer to the above frequency band as the \ltp Measuring
Bandwidth (MBW) in the sequel. Equation~\eref{eq.1} is ten times less
demanding than what is needed for \lisa~\cite{lisascrd}, both in
magnitude and in frequency band, yet it is between two and three
orders of magnitude better than has been achieved or required so
far for space missions~\cite{case}. It has relevant consequences
for future missions, which need high performance drag free, hence
the relevance of \lpf beyond its natural objectives as a \lisa precursor.

\begin{figure}[b]
\centering
\includegraphics[width=7.5cm]{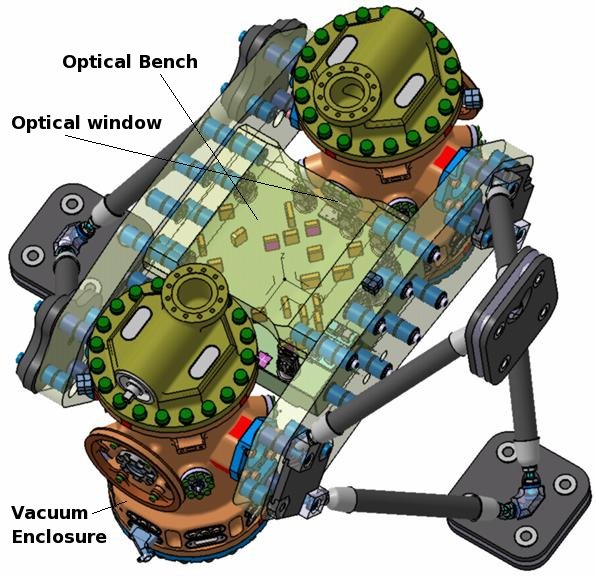}
\caption{Layout of the \ltp Core Assembly, as of 13-October-2006. The
test masses are Au-Pt alloy cubes inside either vacuum enclosure.
\label{fig.1}}
\end{figure}

In order to meet the above requirements, the residual pressure inside
the GRS must be under 10$^{-5}$ Pa, a condition which is classified as
\emph{very high vacuum} by the American Vacuum Society. This implies that their
interior have to be tightly sealed within a \emph{vacuum enclosure} (VE),
and non-mechanical getter pumps installed to ensure a suitably rarefied
environment around the test masses. Perhaps a more obvious option would
have been to communicate the VE directly with the external interplanetary
vacuum, which is much better than 10$^{-5}$ Pa. Recent studies~\cite{asdvacuum}
by the Project Engineering Team have shown that there are serious difficulties
with such option. For example, venting out of residual gas has time scales
exceeding the very LPF mission lifetime, cleanliness control inside the VE
is tighter with the OW, etc.

The general layout is shown in
Figure~\ref{fig.1}. As regards the issues we address in this paper,
attention is drawn to the \emph{Optical Window} (OW), which is the
interface between the test masses and the Optical Bench: laser beams
must bounce off the test masses to monitor their positions by precision
interferometry, hence a transparent window is necessary for the light
to make it to the interior of the VE.

The OW is a plane-parallel plate and is therefore a potential source of
noise: random variations of its optical properties may result in
corresponding optical path fluctuations, which distort the laser light
phase, hence the optical Metrology readout. Great care must be taken when
manufacturing this critical component of the \ltp and, once manufactured,
characterisation of its behaviour duly performed.

The most important agent responsible for OW fluctuations is
\emph{temperature} fluctuations. These cause various degrees of
mechanical stresses across the rim, as well as temperature dependent
index of refraction changes. The former are very difficult to model
with quantitative accuracy, mostly due to lack of precise control of
mounting interface behaviours, but the former can be much better studied
in a stress free environment. This paper is concerned with the experimental
characterisation on ground of prototype OWs, and with the phenomenological
modeling of their response to thermal excitations. This is justified if the
noise fluctuations are smaller than the applied stimuli and if the system
behaves linearly.

The philosophy of the approach is the one typical of the \emph{diagnostics
subsystem}, as described in~\cite{lobo}. This is: apply controlled
temperature signals of high signal-to-noise ratio to the Titanium
flange where the OW is held ---see below---, and measure the
temperature of the former. Measure also the induced phase shifts in
a laser beam which travels through the OW, then try to establish the
transfer function between both magnitudes, i.e., temperature and
phase-shifts. The transfer function thus obtained is assumed to also
be valid in the situation when only noise is present in the flange.
The latter extrapolation hypothesis is the clue to the determination
of the phase noise contributed by temperature fluctuation noise in the
OW, on the basis of temperature measurements.

The success of the proposed empirical approach depends on our ability
to find a transfer function which depends on a (preferably reduced)
number of parameters, which do not change significantly across different
conditions and runs of the experiment. As we shall now show, we have found
that a single pole \textsl{ARMA}\footnote{
Acronym for Auto Regressive Moving Average, roughly the discrete time
series equivalent of a linear differential equation ---see a standard
textbook, e.g.,~\protect\cite{Kay}}.
process describes rather satisfactorily the
relationship we look for. The precise meaning of this concept will be
discussed in detail in the following sections, and the results applied
to evaluate the temperature fluctuation noise in a dedicated experiment.

The paper is organised as follows: in section~\ref{sec.2} we describe
the experiment layout, including hardware and data acquisition details.
Section~\ref{sec.3} is devoted to the data processing, analysis, two
modes of model fitting ---a direct linear regression and a single pole
\textsl{ARMA} model--- and numerical results. In section~\ref{sec.4} we
examine in detail the \textsl{ARMA}(2,1) fit, and derive important
implications for the understanding of the physical processes happening
in the OW. Section~\ref{sec.5} addresses how the previous analysis can
be applied to quantify the contribution of temperature fluctuation noise
to the total phasemeter noise, based on another set of experimental results,
and section~\ref{sec.6} develops an interesting exercise whereby a
continuous time model is suggested as the origin of the discrete time
\textsl{ARMA} fit. Finally, Conclusions and bibliographic references
close the article.

\section{Experiment description}
\label{sec.2}

The current baseline of \lisa \textsl{PathFindfer} and \lisa includes
vacuum tanks containing the test masses which act as end mirrors for
the interferometer. Presence of such tanks, or vacuum enclosures (VE),
force the inclusion of a transmissive element interfacing between the
interior of the VE and the optical bench outside. This optical element is
the \emph{Optical Window} (OW). In this section we describe the laboratory
hardware and conditions of several runs of measurements conducted in AEI
Hannover Laboratory facilities to characterise the thermal behaviour
of the OW.

\begin{figure}[t]
\centering
\includegraphics[width=0.485\columnwidth]{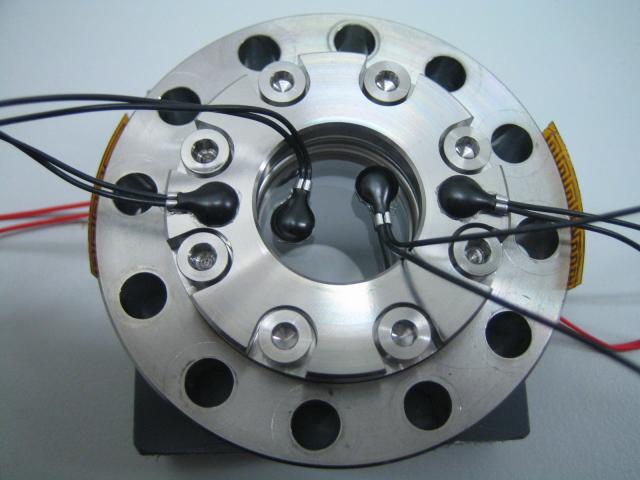}\quad
\includegraphics[width=0.467\columnwidth]{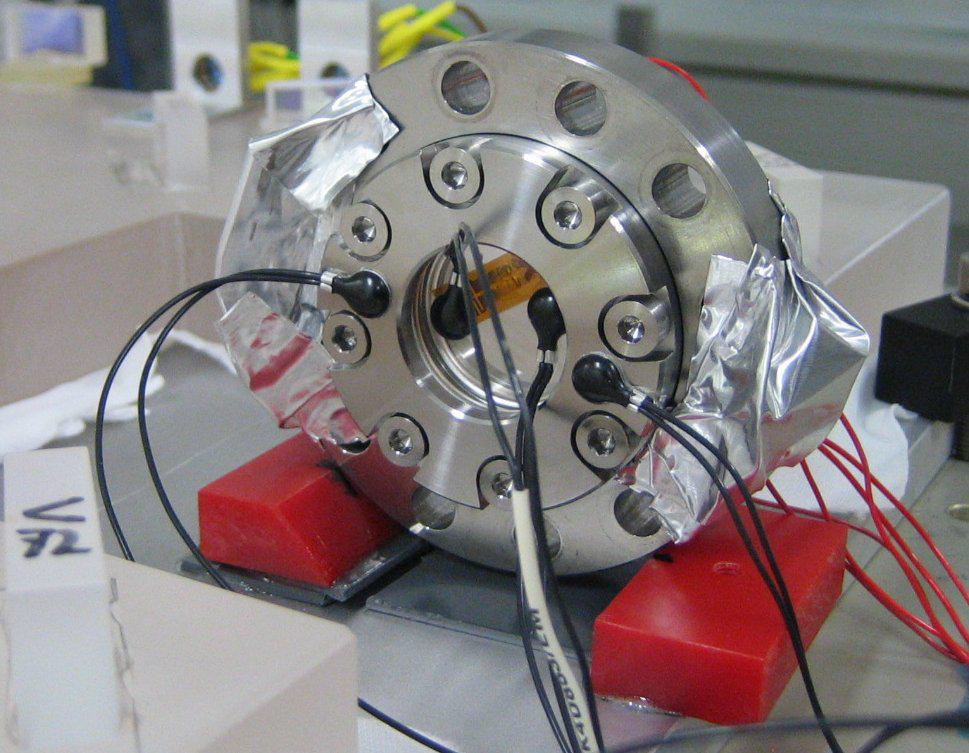}
\caption{The OW (left), with the plane-parallel plate inside the Titanium
flange, heaters on the lateral surface of the latter (pale brown foils)
and two pairs of NTC temperature sensors (black beads with wires). On
the right, mounting of the OW on rails for measurements. Note heaters
are covered with aluminium foils: this is to prevent undesired heating
of other components by heaters' emission of thermal radiation.}
\label{fig:exp}
\end{figure}

In the experiment two different prototype OWs were tested. Both were
manufactured following the same baseline as the one to be applied in the
final \ltp flight model. The main element of the window is a very low
thermal expansion coefficient glass chosen in order to minimise the
variation of the optical pathlength with respect the temperature. The
figure of merit ---quantified by equation \eref{eq.21d} below--- is of
$0.59\times 10^{-6}\,{\rm K^{-1}}$ for our particular choice, the
\textsl{OHARA S-PHM52} ($n\/$\,=\,1.606,
$dn/dT$\,=\,$-5.54\times 10^{-6}\,{\rm K^{-1}}$). This parameter
can reach values as high as $5.31\times 10^{-6}\,{\rm K^{-1}}$ for BK7
or $8.32\times 10^{-6}\,{\rm K^{-1}}$ for fused silica. The glass, 
of 30\,mm diameter and 6\,mm length was clamped between two
Titanium flanges, fastened by means of Titanium bolts, and
sealed by two \emph{helicoflex} rings\footnote{
These are softer metallic rings, e.g.\ Aluminium or Silver.}
to prevent gas leakage in space conditions.

The OW is expected to induce
thermal related noise in the Metrology subsystem. In order to quantify
its contribution to the total noise budget a set of thermal diagnostics
items were attached to the optical window prototypes. They are shown in
figure~\ref{fig:exp}, left panel: two Kapton heaters \textsl{Minco HK5303}
attached to the Titanium flange lateral surface, and four glass encapsulated
thermistors \textsl{Betatherm G10K4D853} attached in pairs to the Titanium
flange and on the athermal glass surface, for precision temperature
measurements. These diagnostics items were all glued to their attachment
points with Pressure Sensitive Adhesive (PSA) tape \textsl{3M-966}, of
similar characteristics to the one to be used in flight. The temperature
sensors on the glass will actually not fly with the \ltp. They will however
provide relevant information to implement real mission data analysis
procedures and methods, for which only the Titanium temperature data
will be available.

During the experiment, the window was leaning vertically on a PVC two-rail
structure ---see figure~\ref{fig:exp}, right panel---, which impeded any
high conductivity thermal contact with the rest of the hardware. Although
not directly affecting the thermo-optical interaction studied here, the
OW will be part of the VE in the real \ltp, thus a higher thermal
conductance is to be expected, and therefore a faster suppression
of thermal gradients is foreseen during mission operations.

\begin{figure}[t]
\centering
\includegraphics[width=0.75\columnwidth]{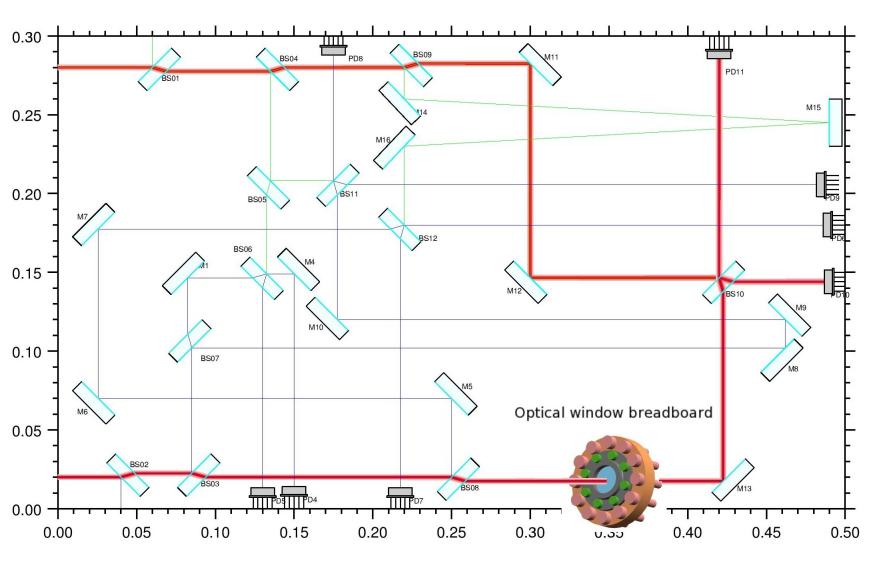}
\caption{Schematic of the interferometric measurement concept.}
\label{fig.3}
\end{figure}

The complete set-up (i.e., the glass plus its mounting structure and the
just mentioned diagnostics items) was inserted as a transmissive element
in a dedicated optical bench, as seen in figure~\ref{fig.3}. The heaters
were covered with aluminium foil to reduce thermal radiation effects
(figure~\ref{fig:exp}, right). For the same reason, the window was
introduced in a copper box leaving only a narrow opening for the
laser beam to go through. As seen in the schematic of figure~\ref{fig.3},
the beam traverses the OW only once. This will not be the case in the real
\ltp, where the laser will go twice through each window, instead, but the one
passage configuration used here simplifies the OW thermal characterisation
without information losses. All the experiments were performed under low
pressure conditions at a $P\simeq 10^{-3}\,\mathrm{Pa}$ vacuum level. 

\begin{figure}[b]
\centering
\includegraphics[width=0.75\columnwidth]{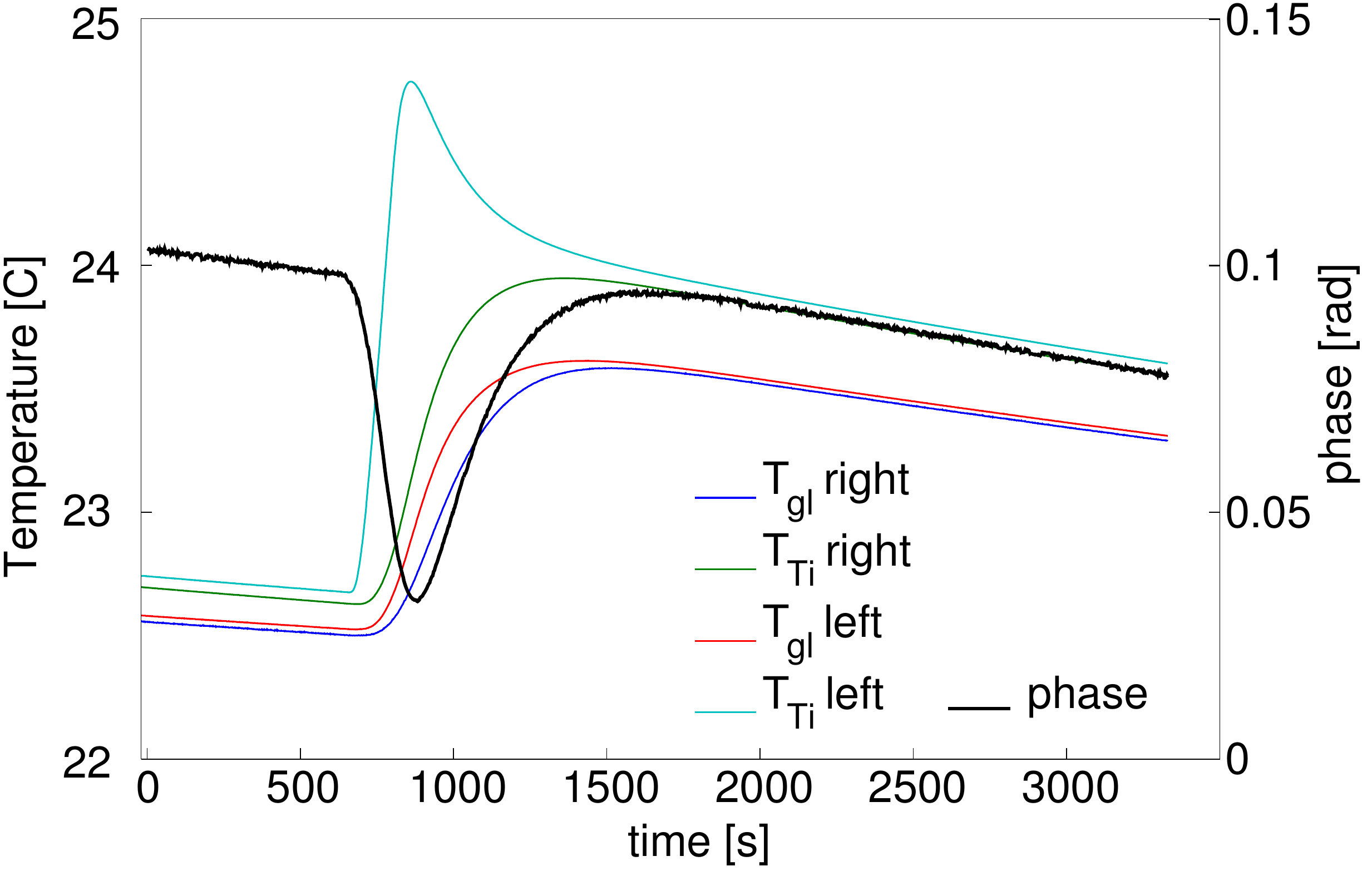}
\caption{Phase (black curve) and various temperature sensors' responses
(coloured curves) to a $2\,\mathrm{Watt}$ heat pulse applied during
$100\,\mathrm{seconds}$. Legend indications correspond to the
thermometers shown in Figure~\protect\ref{fig:exp}.}
\label{fig.4}
\end{figure}

The optical window was subjected to various heat pulses comprising a wide
range of duration length and powers in order to identify suitable
parameters for the thermal test to be performed in-flight. The data
here reported gather 25 experiment runs on two different prototypes,
applying heat pulses from $100\,\mathrm{mW}$ to $2\,\mathrm{W}$ ranging
from $10\,\mathrm{s}$ to $100\,\mathrm{s}$ of application time. All
experiments were performed at room temperature, which falls within
the expected range of working temperatures of the \ltp experiment
during operations, required to be between $10\,^\circ$C and $30\,^\circ$C.
Figure~\ref{fig.4} shows a typical response data plot, with indication
of the temperature sensor readings and the interferometrically registered
phase shifts corresponding to a specific heat signal input ---see the
figure caption for the details.

Two different data acquisition systems were used in the experiment:
the interferometric data were acquired via the \ltp phasemeter
prototype~\cite{gerhar}, whereas the thermal diagnostics data were
acquired using the \ltp front end electronics (FEE)
prototype~\cite{Lobo05,Ramos05}. Both acquisition systems have
previously successfully passed tests of compliance with mission noise budgets.

\section{Fitting the data to models}
\label{sec.3}

The main purpose of this section is to give account of the \emph{measured}
interferometer output data in terms of the also measured temperature data.
While in this experiment both are of course ultimately caused by the
heaters' signal, our interest focuses on the temperature \emph{vs.} phase
relationship, as this is the one we need to quantify the magnitude of
temperature fluctuations noise during science operations in
flight~\cite{lobo}.

To serve this purpose, we adopt model fitting techniques. Two approaches
will be proposed, and discussed in the ensuing section: a direct linear
regression fit of the interferometric data to the temperature read-out
coming from sensors on the Titanium flange and those on the OW
glass itself, and an \textsl{ARMA} model using only temperature readout
from the Titanium temperature sensors. The latter is of particular
interest, since it is not foreseen that temperature sensors be attached
to the glass surface in the real \ltp.

\subsection{Data conditioning}
\label{sec.3-1}

Before we attempt to fit the data to a useful model, some data
pre-processing is required.

The temperature and phase acquisition data systems reside on different
hardware and software, and deliver the respective time series data for
analysis at sample rates which are different as well: temperature data
are sampled at $f_{{\rm s,}T}=0.65\,\mathrm{Hz}$, whereas phase data are
sampled at $f_{{\rm s,}\phi}=32.4\,\mathrm{Hz}$, instead. Downsampling
and resampling thus needs to be applied to the latter in order to make
meaningful sense of data fitting algorithms. To avoid aliasing effects
at downsampling phase, suitable low pass filters are applied. This is
however not enough to have matched sampling times in both time series,
so an additional interpolation algorithm is used for properly matched
resampling.

In addition, each data segment is de-trended prior to model fitting. The
removed trend is evaluated from the first $500$~seconds previous to the
heat input signal begins. This way we get rid of environmental drifts and
spurious trending effects.

\subsection{Direct Linear Regression}
\label{sec.3-2}

A typical phasemeter response when heat pulses are applied to the OW is
shown in Figure~\ref{fig.4}. An essentially instantaneous phase response
is observed in coincidence with thermometers' excitations, which suggests
phase behaviour can be described as a direct, or single-time relationship
between the various temperature readings and associated phase shifts. If
we additionally make the hypothesis that such relationship is
\emph{linear}\footnote{
This is in fact quite accurate, on account of the rather small temperature
and phase variation ranges detected in the experiment.}
then the model is given by
\begin{equation}
 \phi(t) = p_1\,T_{\rm Ti}(t) + p_2\,T_{\rm Glass}(t)
 \label{eq.2}
\end{equation}
where $T_{\rm Ti}(t)$ is the temperature read by the thermometer on the
Titanium flange closest to the activated heater, and $T_{\rm Glass}(t)$
that of a thermometer on the OW glass. The parameters $p_1$ and $p_2$ are
to be estimated by a least squares algorithm, which requires the square error
\begin{equation}
 \epsilon^2 = \sum_{n=1}^{N}\,\left\{\phi[n]- p_1\,T_{\rm Ti}[n]
 -p_2\,T_{\rm Glass}[n]\right\}^2
 \label{eq.3}
\end{equation}
to be the smallest possible for the given data streams. Here, $\phi[n]$
is the $n\/$-th sample of the measured phase, and $T_{\rm Ti}[n]$ and
$T_{\rm Glass}[n]$ the corresponding temperature samples. Thus, for
example,
\begin{equation}
 \phi[n] \equiv \phi(n\Delta t)\ ,\quad
 T_{\rm Ti}[n] \equiv T_{\rm Ti}(n\Delta t)\ ,\quad
 \label{eq.4}
\end{equation}
etc., where the \emph{sampling time} $\Delta t$ has been set to
$\Delta t$\,$\equiv$\,$1/f_{{\rm s,}Temp}$, as discussed in
section~\ref{sec.3-1} above. Finally, $N\/$ is the number
of analysed samples of each read-out.

\begin{figure}[t]
\centering
\includegraphics[width=0.75\columnwidth]{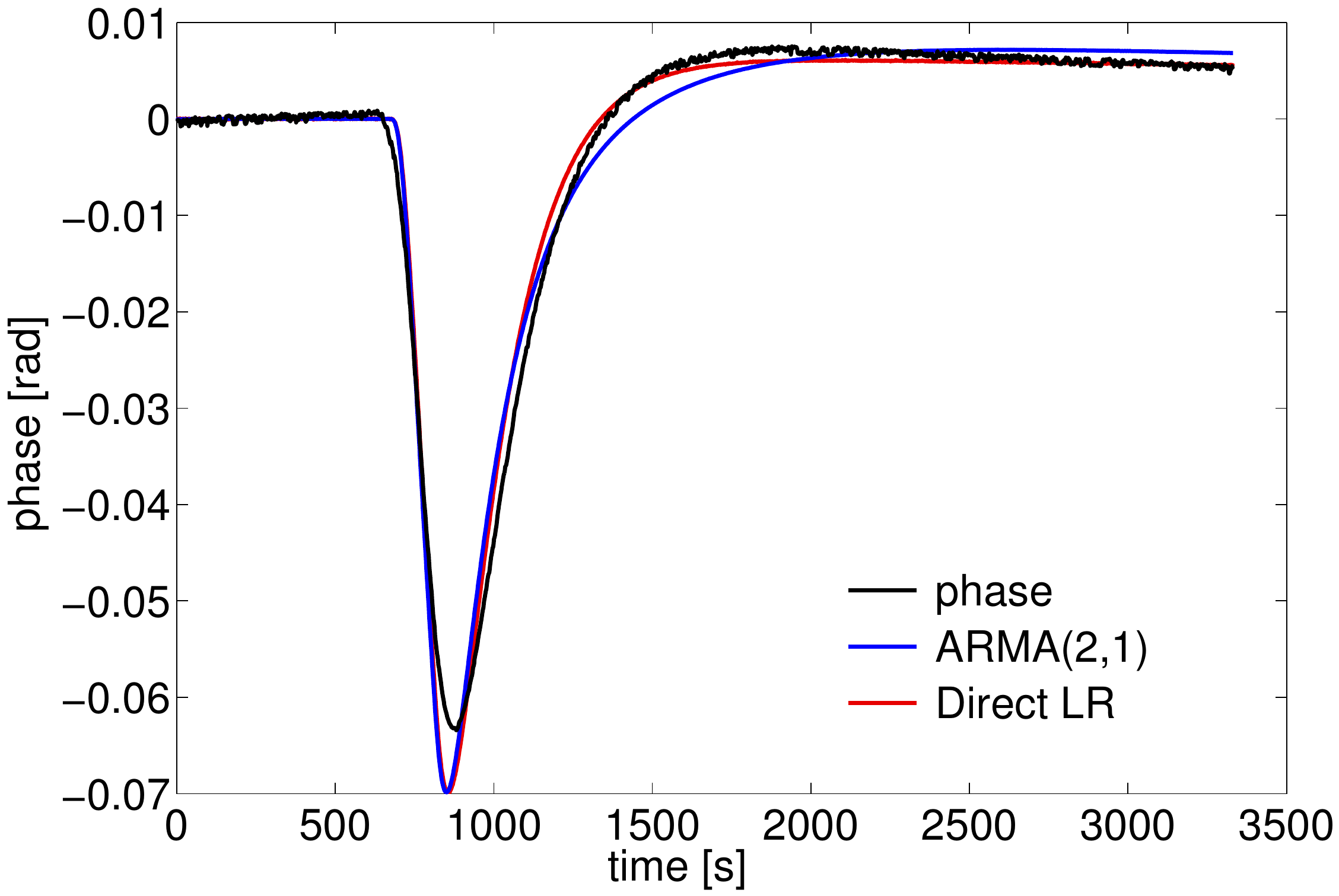}
\caption{Example of fit results for two different models.}
\label{fig:FitEx}
\end{figure}

The conditions of minimum square error are of course given by the two
equations
\begin{equation}
 \frac{\partial\epsilon^2}{\partial p_1} =
 \frac{\partial\epsilon^2}{\partial p_2} = 0
 \label{eq.5}
\end{equation}
which, once solved, give \emph{least squares estimates} $\widehat{p}_1$
and $\widehat{p}_2$ of the parameters $p_1$ and $p_2$, respectively.
An example of this procedure is shown in figure (\ref{fig:FitEx}). We
report on the results of this analysis in section~\ref{sec.3-4}.

\subsection{\textsl{ARMA} model fit}
\label{sec.3-3}

Although the linear regression method shows to perform quite acceptably
well, there is a clear motivation to find a model able to fit the data
independently of the glass temperature readings, since the latter will
not be available in flight.

In this section we take a different approach to fit phase data to Titanium
only temperature readings. We shall now assume that the relationship
between these magnitudes can be expressed by a \emph{dynamical} equation.
More specifically, we make the hypothesis that phase is the output of a
\emph{linear} \textsl{ARMA} process\footnote{
We feel again justified in assuming a \emph{linear} relationship
by the small variation intervals of the magnitudes involved.},
whose input is the temperature of the Titanium, as recorded by
the thermometer next to the activated heater. We express this
by the formula~\cite{Ljung}
\begin{equation}
 \phi[n] = G(q,\bma{\theta})\,T_{\rm Ti}[n] 
 \label{eq.6}
\end{equation}
where \ensuremath{G(q,\bma{\theta})} is a rational expression of the type
\begin{equation}
 G(q,\bma{\theta}) = \frac{\alpha_0 + \alpha_1\,q^{-1} + \cdots +
 \alpha_r\,q^{-r}}{1 + \beta_1\,q^{-1} + \cdots \beta_s\,q^{-s}}
 \label{eq.7}
\end{equation}
with $q\/$ representing the \emph{shift operator}:
\begin{equation}
 q\,x[n] = x[n+1]\ ,\quad q^{-1}\,x[n] = x[n-1]
 \label{eq.8}
\end{equation}
for any discrete series \ensuremath{x[n]}. Finally, \bma{\theta} is an
abbreviation for the vector of $r\/$\,$+$\,$s\/$\,$+$1 \textsl{ARMA}
parameters $\alpha_0,\ldots,\alpha_r,\beta_1,\ldots,\beta_s$.

System identification in this approach is again based on a least squares
criterion, for which a suitably defined square error needs to be defined.
Following~\cite{Ljung}, this is the so called \emph{prediction error}:
\begin{equation}
 \epsilon^2(\bma{\theta}) = \sum_{n=1}^N\,\left\{
 \phi[n] - G(q,\bma{\theta})\,T_{\rm Ti}[n]\right\}^2
 \label{eq.9}
\end{equation}

The estimates \ensuremath{\widehat{\bma{\theta}}} of the parameters
\bma{\theta} are those which cause $\epsilon^2(\bma{\theta})$ to
be minimum. Algorithms to find them are more robust if the additional
hypothesis holds that the residuals
\ensuremath{\left\{\phi[n] - G(q,\bma{\theta})\,T_{\rm Ti}[n]\right\}},
where $\phi[n]$ and $T_{\rm Ti}[n]$ are the actually recorded data, are a
white noise sequence~\cite{Ljung}. Reassuringly, this is quite accurately
true for our data. An example result of the fit is shown in
figure~\ref{fig:FitEx}, blue curve.

\subsection{Numerical results}
\label{sec.3-4}

As stated in section~\ref{sec.2}, up to 25~rounds of measurements were
carried through during the experiment. This is a considerable number
which enables us to check the consistency of the fitting models just
described. The methodology we have adopted is the following: for each
run, we de-trend the data and then fit them to both the Direct Linear
Regression (DLR) and the \textsl{ARMA} models. Parameter estimates are
then filed for further analysis, as we now describe. An observation on
the \textsl{ARMA} fit is however in order before we proceed.

Indeed, in the \textsl{ARMA} fit we also need to make a choice of
\emph{order} of the process, i.e., we need to set the number of
$\alpha$'s and $\beta\/$'s in equation~(\ref{eq.7}). It turns out that
an \textsl{ARMA}(2,1), i.e., two $\alpha$'s and one $\beta\/$, is an
excellent approach, in the sense that differences between model
predictions and actual phase data are kept small to a rather
satisfactory level. Finer tuning can be accomplished adding more
$\alpha$'s and/or $\beta\/$'s, but only at the expense of excessive
parameter estimates' dispersion across different runs. This is highly
undesirable, and hence we confine our model to the \textsl{ARMA}(2,1).

\begin{table}
\caption{Mean values and \emph{rms} variances of parameter estimates}
\label{tab.1}
\centering
\begin{tabular}{l|l}
\multicolumn{1}{c|}{\textbf{DLR}} &
\multicolumn{1}{c}{\textbf{\textsl{ARMA}(2,1)}} \\
\hline \\[-1.5ex]
 $p_1$ = $(-38 \pm 4)\times 10^{-3}$ rad/K &
 $\alpha_0$ = $(39.6 \pm 3)\times 10^{-3}$ rad/K \\
 $p_2$ = $(65 \pm 20)\times 10^{-3}$ rad/K &
 $\alpha_1$ = $(-39.5 \pm 3) \times 10^{-3}$ rad/K \\
 & $\beta_1$ = $-0.996 \pm 0.001$
\end{tabular}
\end{table}

\begin{figure}[b]
\centering
\includegraphics[width=0.48\columnwidth]{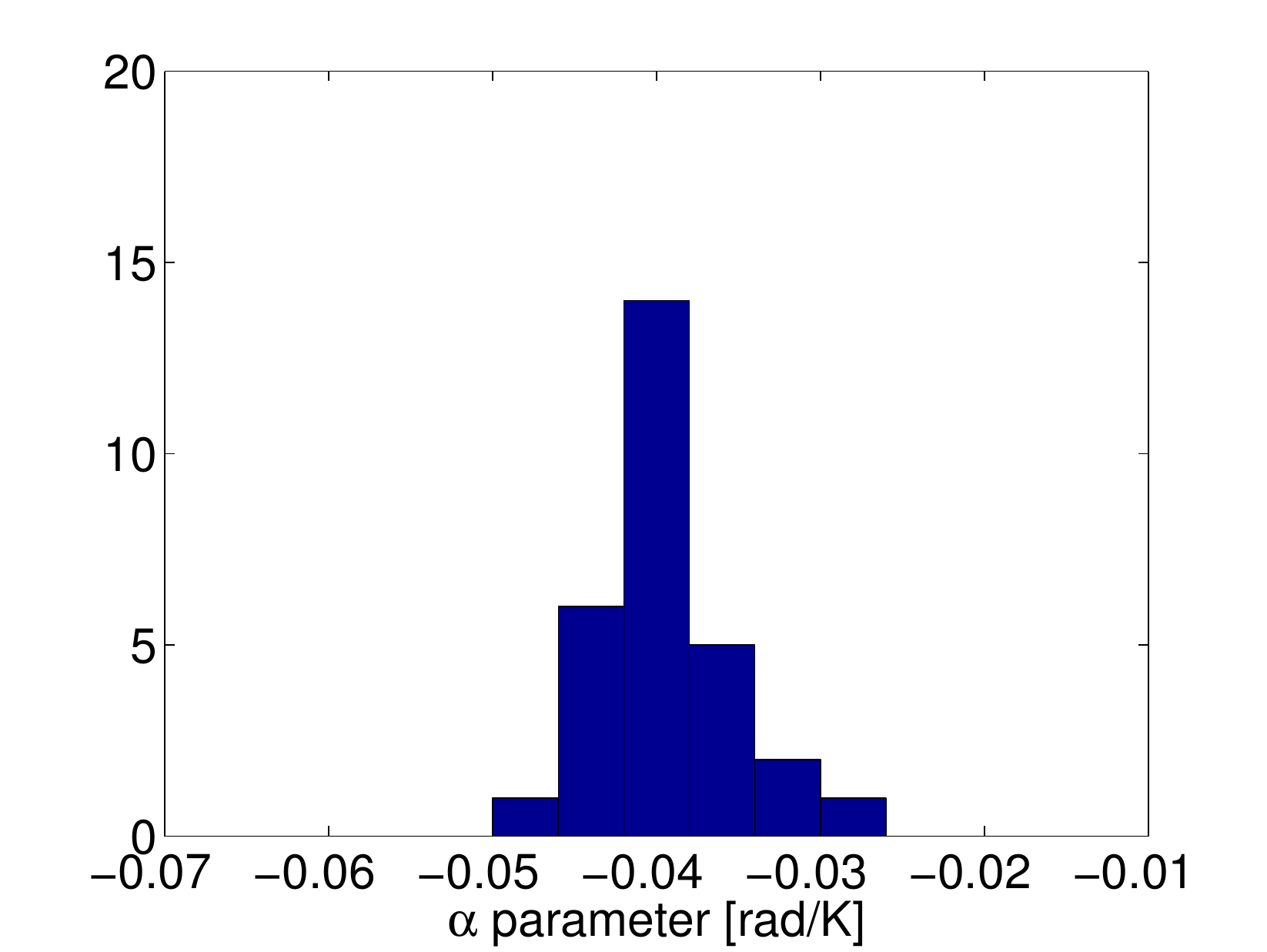}
\includegraphics[width=.48\columnwidth]{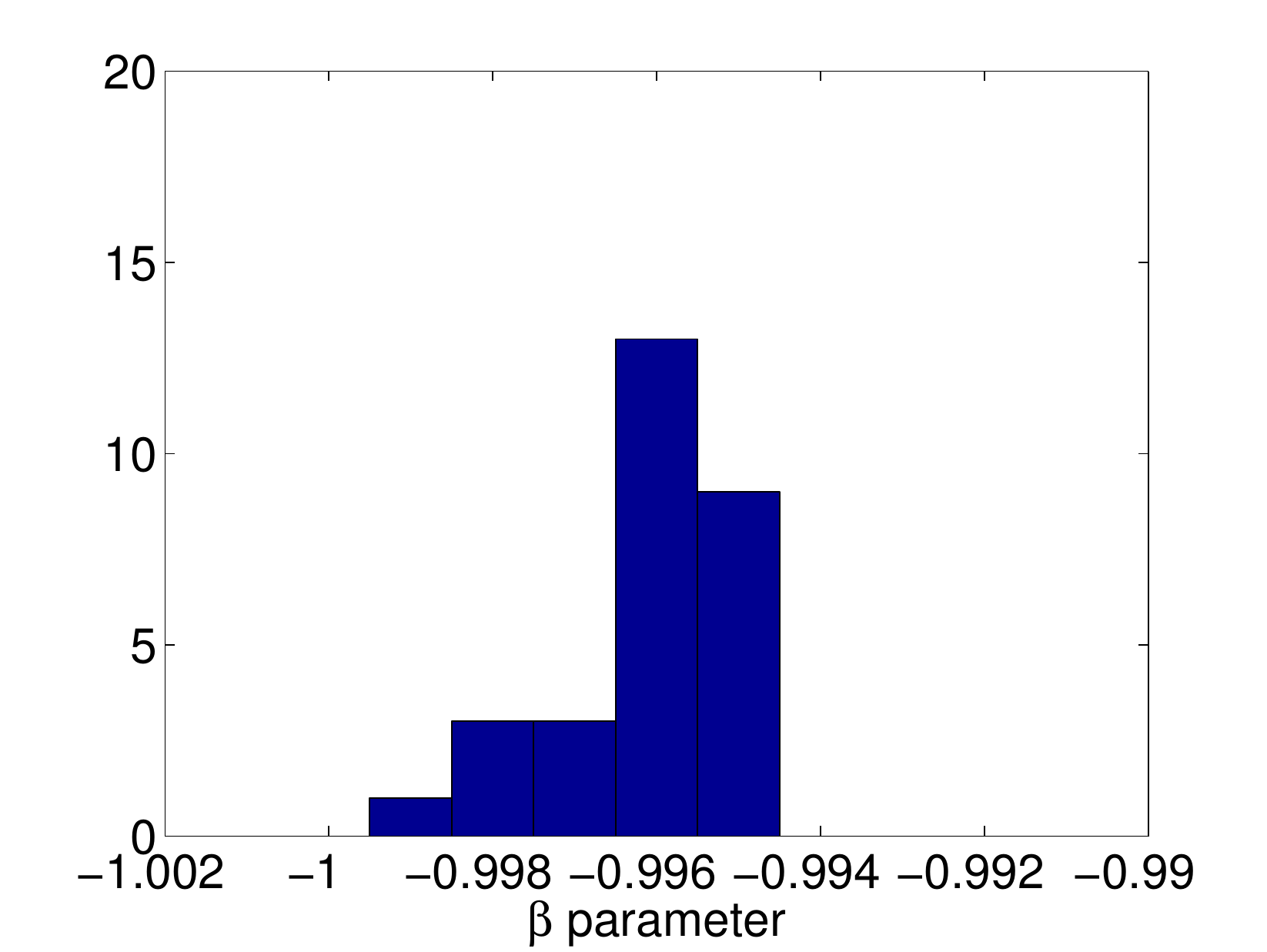}
\includegraphics[width=0.48\columnwidth]{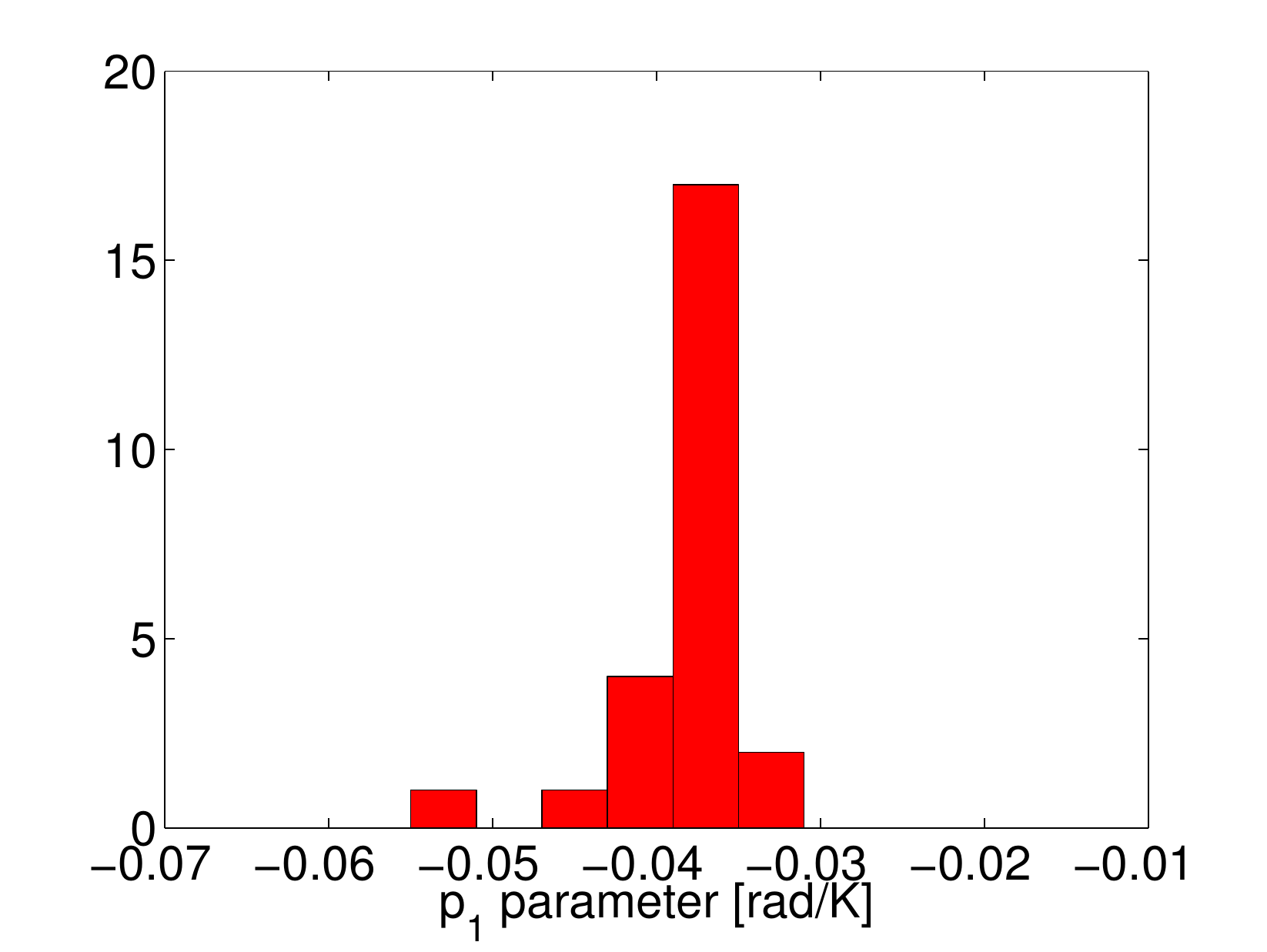}
\includegraphics[width=0.48\columnwidth]{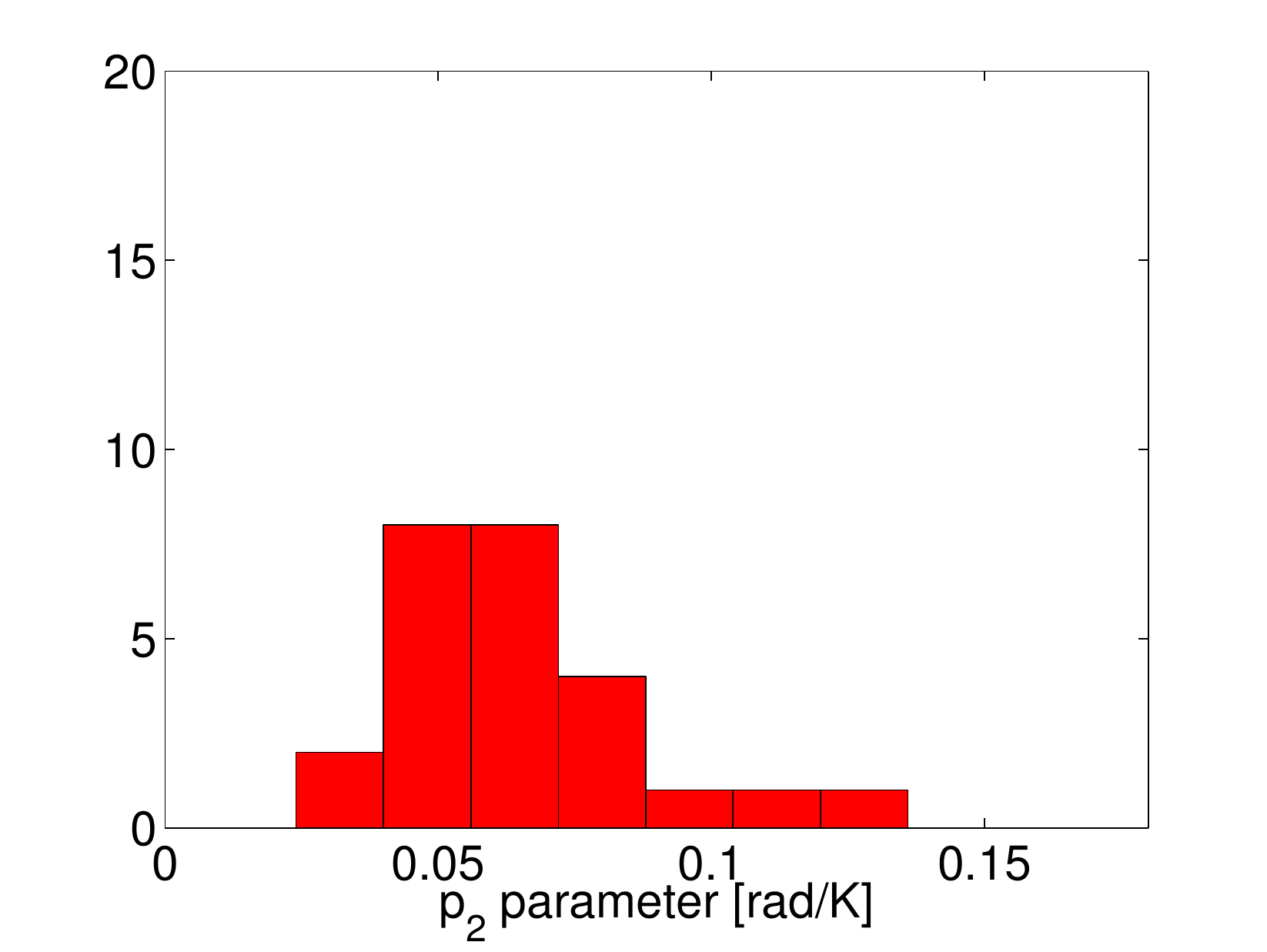}
\caption{Histograms of the estimated parameters for the two fitting models
 described in the text. \emph{Top:} ARMA(2,1), with
 $\alpha$\,$\equiv$\,$\alpha_0$ and $\beta$\,$\equiv$\,$\beta_1$.
 \emph{Bottom:} DLR.}
\label{fig:hist}
\end{figure}

Table~\ref{tab.1} summarises the results of the analysis, and
figure~\ref{fig:hist} shows the binned distribution of the parameter
estimates across the 25 runs. An outstanding charac\-teristic of the fit is
that the two \textsl{MA} coefficients very accurately verify the numerical
relationship $\alpha_1$\,=\,$-\alpha_0$. The model thus relates the
output phase data to the time \emph{derivative} of the Titanium
temperature ---we come back to this in section~\ref{sec.4}.

There are a few important aspects of these results which are worth
stressing:

\begin{itemize}
\setlength{\itemsep}{0 ex}   
 \item $\alpha_1$\,$\simeq$\,$-\alpha_0$, although the difference between
       their values is much less than their variances.
 \item Well within tolerance, $-\alpha_1$\,=\,$\alpha_0$\,=\,$p_1$.
 \item $\beta_1$ is strongly peaked at a nominal value, with only 0.1\,\%
       relative tolerance.
 \item $p_2$ is much more disperse, with almost 30\,\% variability.
\end{itemize}

\section{The \textsl{ARMA}(2,1) transfer function}
\label{sec.4}

In view of the above remarks, it is expedient to rewrite the
\textsl{ARMA}(2,1) model in terms of the following parameters:
\begin{equation}
 \alpha\equiv -\alpha_1\ ,\quad \delta\equiv\alpha_0+\alpha_1\ ,\quad
 \beta\equiv\beta_1
 \label{eq.10}
\end{equation}

Hence,
\begin{equation}
 G(z,\alpha,\beta,\delta) = \alpha\,\frac{1-z^{-1}}{1+\beta z^{-1}} +
 \frac{\delta}{1+\beta z^{-1}}
 \label{eq.12}
\end{equation}
is the $z\/$-transform of the process transfer function ---we have
replaced the shift operator $q\/$ by the complex variable $z\/$~\cite{Kay}.
It is also expedient to emphasise the structure of this formula as follows:
\begin{equation}
 G(z,\alpha,\beta,\delta) = \alpha\,G_{\rm HP}(z,\beta) +
 \delta\,G_{\rm LP}(z,\beta)
 \label{eq.121}
\end{equation}
with
\begin{equation}
 G_{\rm HP}(z,\beta)\equiv\frac{1-z^{-1}}{1+\beta z^{-1}}\ ,\quad
 G_{\rm LP}(z,\beta)\equiv\frac{1}{1+\beta z^{-1}}
 \label{eq.122}
\end{equation}

We now find \emph{discrete Fourier transforms} (DFT) by the substitution
\begin{equation}
 z = \exp(i\omega\Delta t)
 \label{eq.125}
\end{equation}
where $\Delta t\/$ is the sampling time of the time series data. The
following obtains:
\begin{eqnarray}
 |\widetilde{G}_{\rm HP}(\omega,\beta)|^2 & = &
 \frac{4\sin^2(\omega\Delta t/2)}{1+2\beta\,\cos(\omega\Delta t)
 + \beta^2} \label{eq.125a} \\[1ex]
 |\widetilde{G}_{\rm LP}(\omega,\beta)|^2 & = &
 \frac{1}{1+2\beta\,\cos(\omega\Delta t) + \beta^2} \label{eq.125b} \\[1ex]
 |\widetilde{G}(\omega,\alpha,\beta,\delta)|^2 & = & \frac{\delta^2 +
 4\,\alpha(\alpha+\delta)\,\sin^2(\omega\Delta t/2)}{1 +
 2\beta\,\cos(\omega\Delta t) + \beta^2}
 \label{eq.125c}
\end{eqnarray}

We thus see that the transfer function $G\/$ is the sum of a \emph{high pass}
term, $G_{\rm HP}$, and a \emph{low pass} term, $G_{\rm LP}$. The effect of
the latter is naturally dominant at low frequencies, while the high pass
term dominates at high frequencies. The concepts of \emph{low} and
\emph{high} frequencies can be made precise by means of some intermediate
frequency $f_{\rm b}$ where the gains of $G_{\rm HP}$ and $G_{\rm LP}$ are
equal. This is easily calculated:
\begin{equation}
 f_{\rm b} \simeq \left|\frac\delta\alpha\right|\,\frac{1}{2\pi\Delta t}
 \label{eq.126}
\end{equation}
and has a numerical value of $f_{\rm b}$\,$\simeq$\,0.2\,mHz, which means
the high pass effect dominates throughout the \ltp bandwidth. We may not
however neglect the relevance of the low pass at lower frequencies, as it
contributes extremely valuable information for \lisa.

\begin{figure}
\centering
\includegraphics[width=0.98\columnwidth]{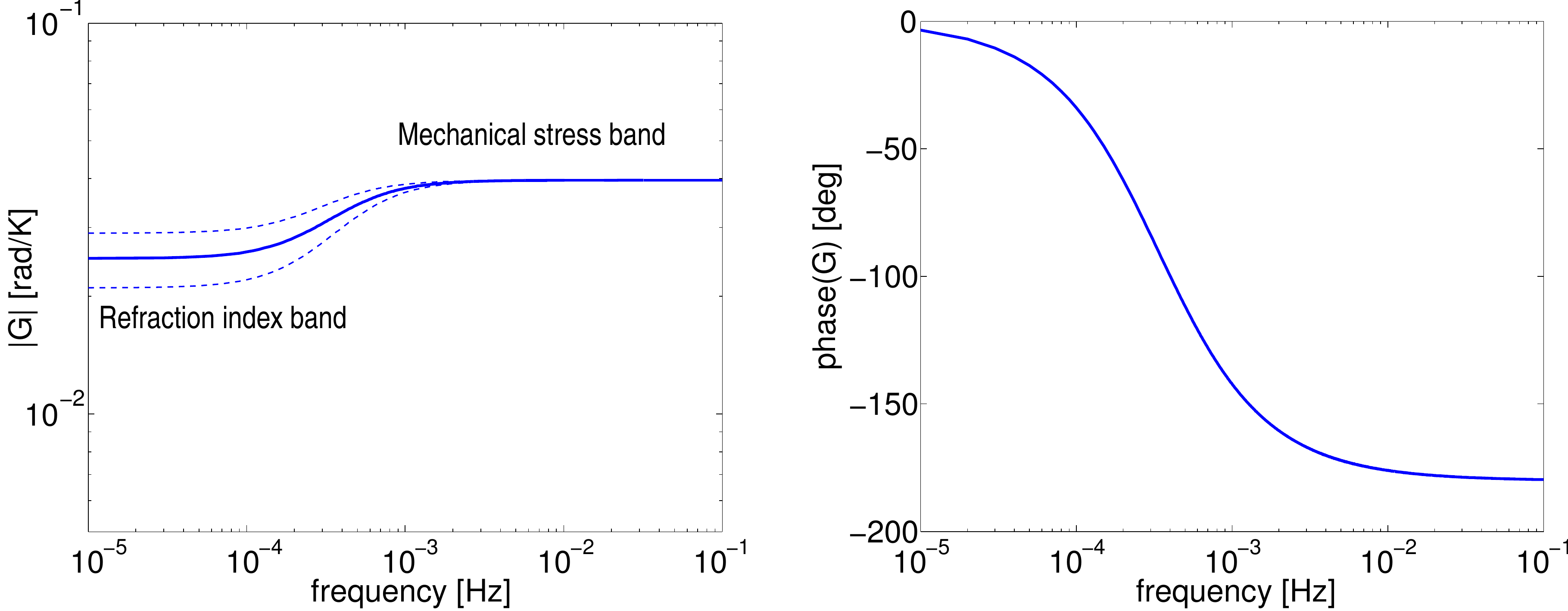}
\caption{Bode diagram for the optical window \textsl{ARMA}(2,1)
transfer function estimate using the values from Table \ref{tab.1}. 
Left panel: its modulus; dashed lines show the $1\,\sigma$ confidence
region. Note that this confidence region has been extrapolated
below $\sim$0.4~mHz, as actual experimental data were actually not
available in that band. Right panel: phase of the transfer function.}
\label{fig:trans}
\end{figure}

A Bode diagram representation for the transfer functions is shown in
figure~\ref{fig:trans}. The filter modulus is characterised by a plateau
of $|\widetilde{G}|\sim 40\times10^{-3}\,\mathrm{rad/K}$ across the entire
\ltp measuring bandwidth. Temperature fluctuations at frequencies below
this bandwidth are also suppressed but following a different behaviour,
related to the low frequency response of the optical window. The Figure
also shows the phase behaviour of the filter, and indicates increasing
delays for high frequency perturbations.

The DC gain of the filter is
\begin{equation}
 |\widetilde{G}(\omega=0,\alpha,\beta,\delta)| = \frac{\delta}{1+\beta}
 \label{eq.21b}
\end{equation}

If the estimated parameters are substituted in this expression then the
following is obtained:
\begin{equation}
 |\widetilde{G}(\omega=0,\alpha,\beta,\delta)| =
 (25\pm 4) \times 10^{-3}\,{\rm rad/K}
 \label{eq.21c}
\end{equation}

We now try to produce some insight into the physical meaning of the
just discussed analysis facts.

\subsection{Physics of the \textsl{ARMA} process}
\label{sec.4-1}

Two different kinds of thermal effects have been identified as sources
of changes in the optical path-length of a light beam traversing a
plane-parallel piece of glass:

\begin{enumerate}
\setlength{\itemsep}{0 ex}   
\renewcommand{\labelenumi}{\sf\roman{enumi}.}
 \item Temperature dependent changes of the refraction index
 \item Mechanical stress induced changes of the refraction index
\end{enumerate}

We briefly describe below how these effects can be approximately evaluated.

\subsubsection{Temperature dependent changes of the refraction index}
\label{sec.4-1.1}

The first effect, which is found under stress free conditions, is quantified
by the formula~\cite{Marinphd}
\begin{equation}
 \left.\frac{d\phi}{dT}\right|_{\rm free} = 2\pi\frac{L}{\lambda}\,\left[
 \frac{dn}{dT} + (n-1)\,\alpha_E\right]
 \label{eq.21d}
\end{equation}
where $\phi$ is the phase shift suffered by a beam of light traversing
a glass slab of thickness $L\/$ and (nominal) index of refraction $n\/$;
$\lambda$ is the wavelength of the used light, and $\alpha_E\/$~is the linear
thermal expansion factor of the glass, $\alpha_E$\,=\,$L^{-1}\,dL/dT$.

The $d\phi/dT|_{\rm free}\/$ effect is most prominent at very low
frequencies and DC. The reason is that it happens even if the temperature
of the glass is homogeneous, and without mechanical stresses. It has been
measured on \emph{naked} glass samples in the laboratory, free of any
pressure or tension, with the result that it is 25\,mrad/K~\cite{amaldi6},
a figure very well matching the one given by equation~\eref{eq.21c}. One
however needs to consider that the latter was obtained from data of a
real window, i.e., including metal flange. This consequently means that
the stress contribution $d\phi/d\sigma$ must be comparatively small at
very low frequencies.

The same result is endorsed by another independent
evidence. If data-sheet properties of the \textsl{OHARA S-PHM52}
glass used in the experiment are used to calculate the thermal related
path-length variations in the optical window glass due to changes in the
refractive index, the result is that $d\phi/dT|_{\rm free}\/$ is
$\sim$21\,mrad/K, again in good agreement with equation~\eref{eq.21c}.

\subsubsection{Mechanical stress induced changes of the refraction index}
\label{sec.4-1.2}

This second effect is relevant to our experiment because the glass, clamped
by Titanium flanges to the ISH structure, is under stress due to
differing thermal expansion coefficients in glass and metal. Mechanical
stress also induces pathlength changes which are difficult to model.
From the datasheet, the only parameter provided by the manufacturer which can
be used to quantify these interactions is the \emph{photoelastic coefficient},
$\beta$. However, it must be noticed that $\beta$ does not describe the
change in the refraction index due to stress, $dn/d\sigma$, but the
appearence of birefringence due to stress, i.e., the change of the 
velocity of light along different axes of the material. Although
not directly related, both parameters range in the same order of
magnitude~\cite{Marinphd}, and we shall thus use the photoelastic
coefficient here for our order of magnitude estimate, described
in the following.

Under this simplyfing assumption, the photoelastic coefficient can be
related to a pathlength variation by
\begin{equation}
 \Delta s_{\rm stress} = \beta\,\sigma\, d
 \label{eq.21e}
\end{equation}
where $\beta$\,=\,10$^{-5}$\,nm\,cm$^{-1}$\,Pa$^{-1}$, $d\/$ is the glass
thickness ($d\/$\,=\,0.6\,cm for the Optical Window), and $\sigma$ is the
applied stress, having dimensions of pressure.

\begin{figure}
\centering
\includegraphics[width=0.7\columnwidth]{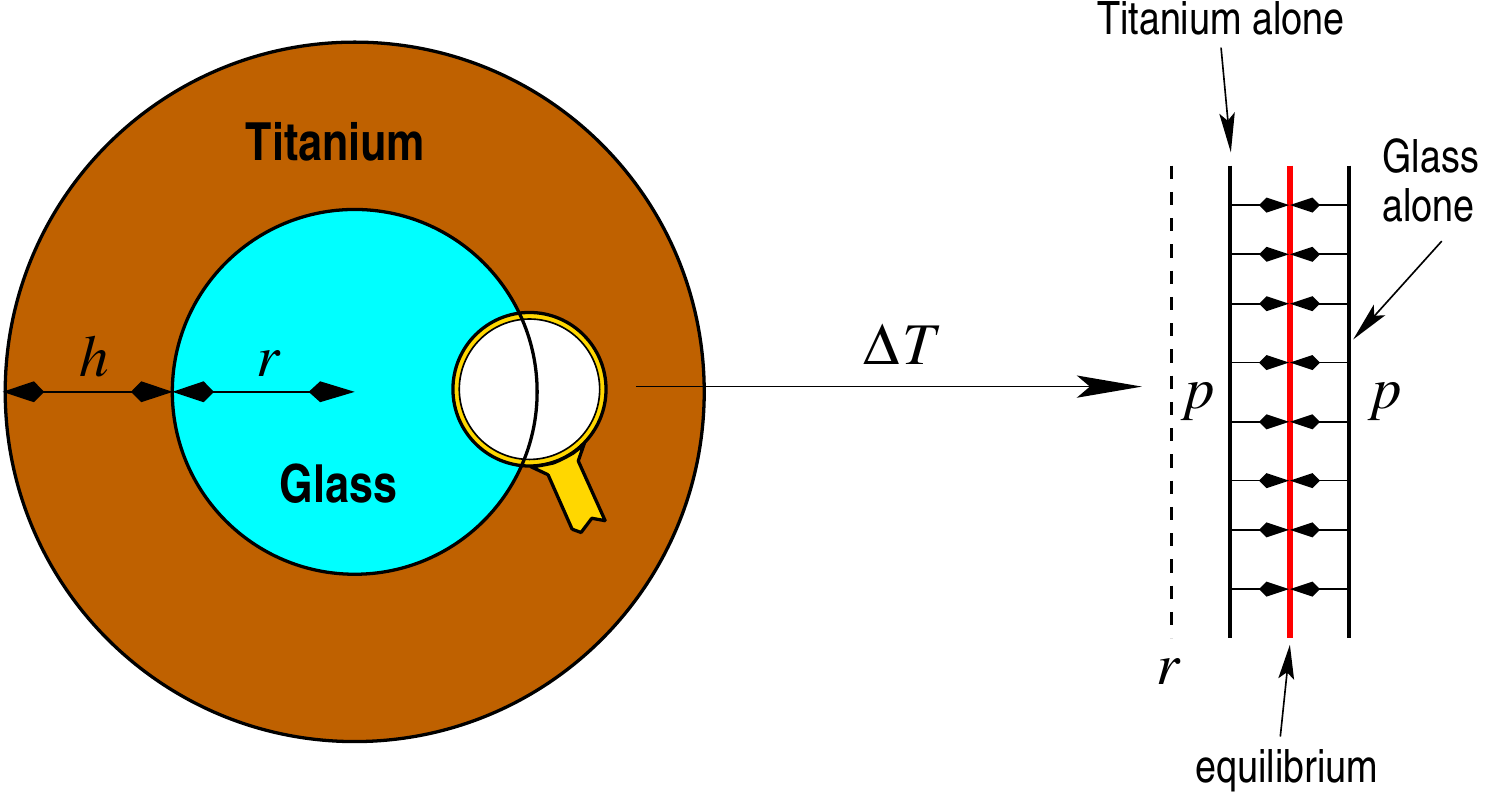}
\caption{Schematics of the dilatation of the OW glass and the clamping
titanium flange: the right part zooms in the profiles acquired by the
interface (red) when the temperature of the block increases by $\Delta T$.
Should either the Glass or the Titanium be let expand freely, the boundaries
would be placed as also represented. Dashed line is the interface position
before heating.}
\label{fig.8}
\end{figure}

In this case, the stress on the glass is due to \emph{differential} thermal
dilatation of the Titanium flange and the OW glass itself. The situation is
illustrated graphically in Figure~\ref{fig.8}. Because the coefficient of
thermal expansion of the glass is larger than that of the Titanium
flange embracing it, the latter expands less when submitted to the
same temperature rise, and hence the glass is compressed radially along
the rim. The opposite happens if the temperature decreases, i.e., the
glass is in this case stretched outwards by the radial pull of the
Titanium. The contraction/expansion forces acting on glass and Titanium
reach an equilibrium state which determines the radii of the
contracted/expanded pieces of Titanium and Glass. The equilibrium position
thus happens when
\begin{equation}
 \left[\delta \rho_T + \delta \rho_\sigma \right]_{\rm Ti} =
 \left[\delta \rho_T - \delta \rho_\sigma \right]_{\rm Glass}
 \label{eq.21ea}
\end{equation}
where $\delta \rho_T$ and $\delta \rho_\sigma$ refers to changes in radius
caused by temperature changes and by stresses, respectively. The above
formula holds even if temperature changes in Titanium and glass are
unequal. On the other hand, we are not considering in our
description possible effects coming from the \emph{helicoflex} ring between
the Titanium and the glass. As stated above, we are here trying to
get an order of magnitude of the effect based on a simplified mechanical
model, and interface effects are thus not included.

The contributions appearing in equation \eref{eq.21ea} are
given by~\cite{timos}
\begin{equation}
 \delta \rho_T  = \rho \,\alpha\Delta T\quad {\rm and}\quad
 \delta \rho_\sigma  = \frac{p\,\rho^2}{\ell \, E}
 \label{eq.21eb}
\end{equation}
where $\rho$ is the radius of the interface between Titanium and the glass,
$\ell$ stands for the width of the body, $E\/$ is the Young
modulus, $\alpha$ the thermal expansion coefficient and $p\/$ the lateral
pressure. Combining equations~\eref{eq.21ea} and~\eref{eq.21eb}, and
following the notation of figure~\ref{fig.8}, we find the lateral
pressure on the glass:
\begin{equation}
 p = \frac{\alpha_{\rm Ti}\Delta T_{\rm Ti} -
 \alpha_{\rm Glass}\Delta T_{\rm Glass}}{(r/h)\,E_{\rm Ti}^{-1}
 + E_{\rm Glass}^{-1}}
 \label{eq.21ec}
\end{equation}

The strain on the glass lateral surface is given by
$\sigma_{\rm Glass}$\,=\,$prd/(r\,d)$\,=\,$p$, where $d\/$
is the thickness of the window glass ---see~\cite{timos}. Hence,
\begin{equation}
 \sigma_{\rm Glass} =\frac{\alpha_{\rm Ti}\Delta T_{\rm Ti} -
 \alpha_{\rm Glass}\Delta T_{\rm Glass}}{E_{\rm Ti}^{-1}
 + (h/r)\,E_{\rm Glass}^{-1}}
 \label{eq.21ed}
\end{equation}

We can consider two different regimes here: the \emph{low frequency} (LF)
regime and the \emph{high frequency} (HF) regime. The first corresponds
to long duration heat pulses applied on the Titanium flange, actually long
enough that the temperatures of the glass and the Titanium equal each other,
or $\Delta T_{\rm Ti}$\,=\,$\Delta T_{\rm Glass}$\,$\equiv$\,$\Delta T$. In
this case
\begin{equation}
 \sigma_{\rm Glass} = \frac{\alpha_{\rm Ti} - \alpha_{\rm Glass}}
 {E_{\rm Ti}^{-1} + (h/r)\,E_{\rm Glass}^{-1}}\,\Delta T\ ,\quad
 {\rm low\ frequency}
 \label{eq.21ee}
\end{equation}
 
On the other hand, if short heat pulses are applied on the Titanium then
the glass does not have time to respond, and in this we can assume
$\Delta T_{\rm Glass}$\,=\,0 and $\Delta T_{\rm Ti}$\,$\equiv$\,$\Delta T$.
Thus,
\begin{equation}
 \sigma_{\rm Glass} = \frac{\alpha_{\rm Ti}}
 {E_{\rm Ti}^{-1} + (h/r)\,E_{\rm Glass}^{-1}}\,\Delta T\ ,\quad
 {\rm high\ frequency}
 \label{eq.21ef}
\end{equation}

We can use the above formulas in combination with~\eref{eq.21e} to obtain
\begin{equation}
 \left.\frac{d\phi}{dT}\right|_{\rm Stress} = \left\{
 \begin{array}{ll}
 \beta\displaystyle\frac{2\pi d}{\lambda_{\rm laser}}\,
 \frac{\alpha_{\rm Ti} - \alpha_{\rm Glass}}
 {E_{\rm Ti}^{-1} + (h/r)\,E_{\rm Glass}^{-1}}\ , &
 \quad {\rm low\ frequency} \\[1.3em]
 \beta\displaystyle\frac{2\pi d}{\lambda_{\rm laser}}\,
 \frac{\alpha_{\rm Ti}}
 {E_{\rm Ti}^{-1} + (h/r)\,E_{\rm Glass}^{-1}}\ , &
 \quad {\rm high\ frequency} \end{array} \right.
 \label{eq.21eg}
\end{equation}

It is recalled that $\Delta\phi$\,=\,2$\pi\Delta s/\lambda_{\rm laser}$,
where $\lambda_{\rm laser}$ is the laser wavelength. We put numbers here:

\medskip

\begin{center}
\begin{tabular}{l}
$\beta = 10^{-3}\times 10^{-9}$ Pa$^{-1}$ \\
$d = 6\times 10^{-3}$ m \\
$\lambda_{\rm laser} = 1.064\times 10^{-6}$ m \\[1ex]
$\alpha_{\rm Ti} = 8.6\times 10^{-6}$ K$^{-1}$ \\
$E_{\rm Ti} = 11.6\times 10^{10}$ N m$^{-2}$ \\
$h = 0.02$ m \\[1ex]
$\alpha_{\rm Glass} = 10\times 10^{-6}$ K$^{-1}$ \\
$E_{\rm Glass} = 7.15\times 10^{10}$ N m$^{-2}$ \\
$r = 0.015$ m
\end{tabular}
\end{center}

\medskip

to obtain
\begin{equation}
  \left.\frac{d\phi}{dT}\right|_{\rm Stress} = \left\{
 \begin{array}{ll}
 2.5\times 10^{-3}\ {\rm rad\ K}^{-1}\ , & \quad {\rm low\ frequency} \\[1ex]
 15\times 10^{-3}\ {\rm rad\ K}^{-1}\ , & \quad {\rm high\ frequency}
 \end{array} \right.
 \label{eq.21eh}
\end{equation}

\subsubsection{Discussion of the results}
\label{sec.4-1.3}

The total thermal effect is the sum of the above two effects, i.e., optical
pathlength changes induced by pure thermal expansion and by mechanical
stress. The former gives a value of $21\times 10^{-3}\ {\rm rad\ K}^{-1}$
throughout the frequency band, as extracted from datasheet values --- see
section~\ref{sec.4-1.1} above. We can thus summarise the results as shown
in Table~\ref{tab.2}:

\begin{table}[h!]
 \centering
 \caption{$d\phi/dT$, units in mrad K$^{-1}$.}
 \label{tab.2}
 \begin{tabular}{ccc}
 & \textsl{ARMA} & Analytic \\
 \hline
 LF range & 25 $\pm$ 4 & 23.5 \\
 HF range & 40 & 36 \\
 Nude glass & --- & 21
 \end{tabular}
\end{table}

The agreement between the results produced by our simplified model
and the \textsl{ARMA} fit is quite good. Even though the model is not fully
comprehensive of all the physical effects happening in the OW, it can
be considered rather satisfactory from a purely empirical point of view,
hence very useful for practical purposes. Work is currently in progress
for a more thorough approach, and we shall report on new results in due
course.

We conclude from this discussion that the low pass component of the transfer
function is almost exlusively related to the $d\phi/dT|_{\rm free}$ effect,
while the stress effects only show up significantly in the higher frequency
band. This makes sense, as stresses applied along the glass rim quickly
propagate inwards throughout the glass piece.

Although the \ltp spectrum is only above 1\,mHz, an analysis at
frequencies below this one, down to 10$^{-4}$\,Hz and even further, must
be considered of high interest, as the latter frequency band will be
important for \lisa. The experimental data reported in this paper can be
improved to access the lower \lisa band, since they typically consist in
one hour long runs. This is a strong suggestion for the \ltp experiment~plan.

\section{Noise projection}
\label{sec.5}

One of the main scientific objectives of the diagnostics system in the \ltp
is to measure identified environmental disturbances~\cite{toplev}, and to
provide the data and analysis tools to estimate the contribution of those
disturbances to the overall mission noise budget, equation~\eref{eq.1}.
In practice this means the \ltp Data and Diagnostics Subsystem (DDS) must
be able to provide suitable \emph{transfer functions} to convert measured
disturbance noise into test mass acceleration noise. This section is
devoted to describe this procedure in the case of temperature fluctuation
noise in the OW, and to show how it works in an on-ground laboratory
experiment ---to be extrapolated to a space-borne~one.

We will use the results derived in the previous analysis to obtain an
estimation for the thermal contribution to the interferometer performance.
We shall naturally limit ourselves to \textsl{ARMA} model, since it is the
one making sense for real mission purposes, as already discussed.

The basic idea is that the OW transfer function, as determined from
high SNR system response, also applies when there is only (weaker)
noise in the window~\cite{lobo}. For this we shall use the one in
equation~\eref{eq.12}, i.e.,
\begin{equation}
 G(z,\alpha,\beta,\delta) = \alpha\,\frac{1-z^{-1}}{1+\beta z^{-1}}
 + \frac{\delta}{1+\beta z^{-1}}
 \label{eq.20}
\end{equation}

We now show which procedures must be applied to address the problem of
finding the contribution of temperature fluctuations noise in the OW to
the total OW noise. To this end we consider data of temperature and phase
noise generated in a different experiment, and apply to it the methodology
just sketched.

\begin{figure}[t]
\centering
\includegraphics[width=.35\columnwidth,angle=-90]{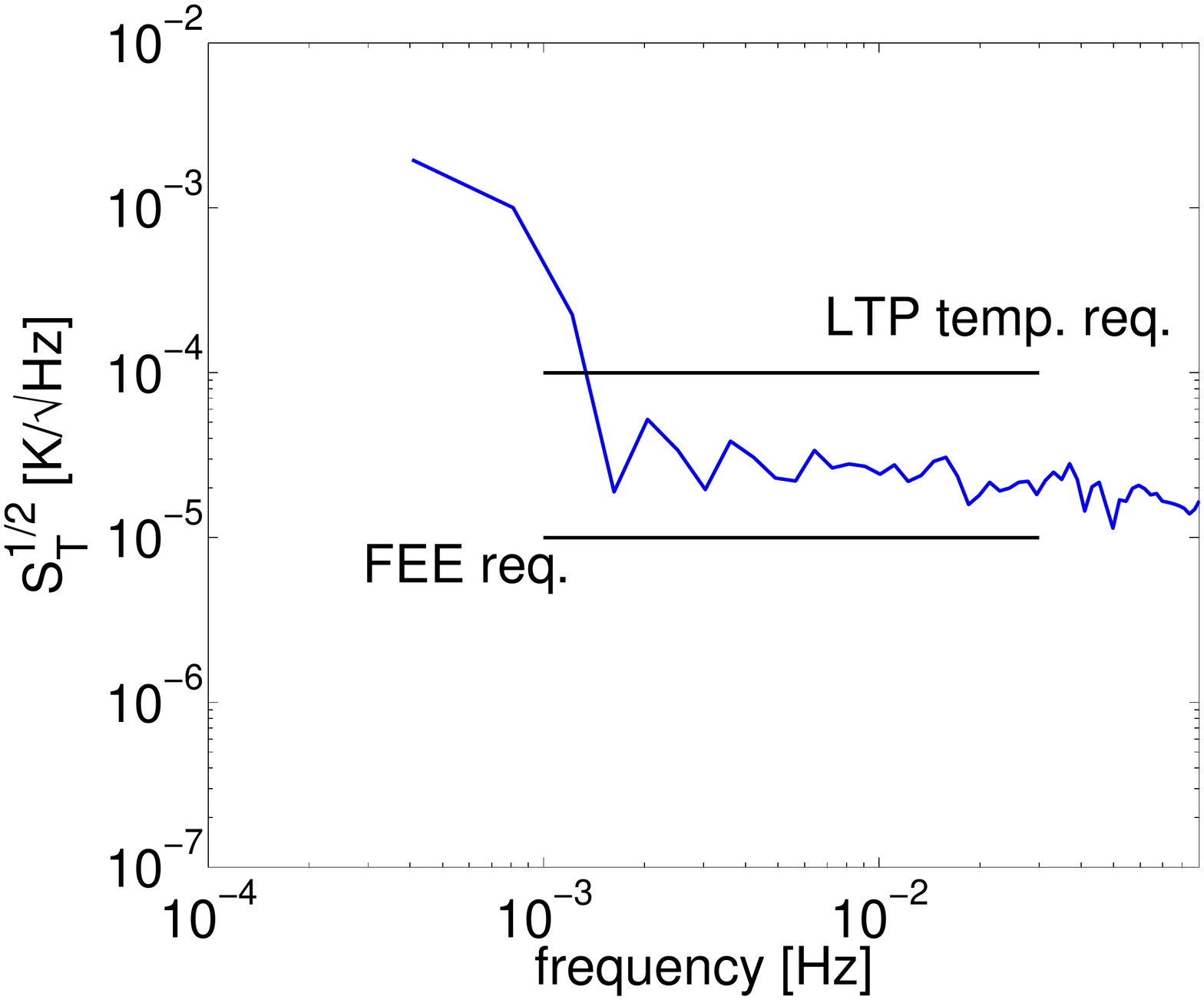}
\includegraphics[width=.35\columnwidth,angle=-90]{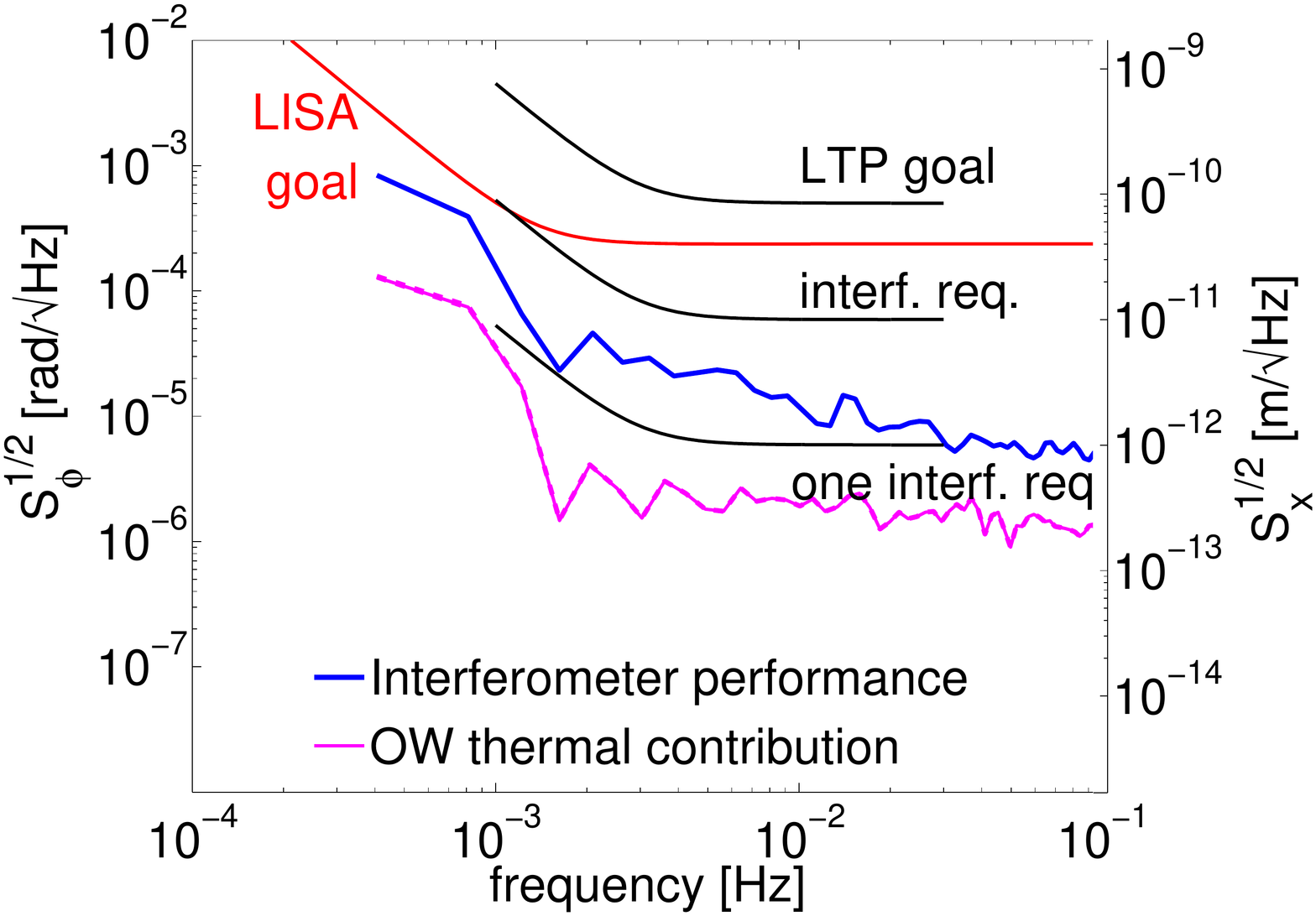}
\caption{Temperature fluctuations spectrum (left) measured during the
performance experiment compared to the \ltp temperature requirement and the
lower limit set by the front end electronics sensitivity limit. Phase
fluctuations (blue curve, right panel) during the same run are compared
to the optical window thermal contribution (magenta curve, right panel),
as derived with the \textsl{ARMA} transfer functions given by
equations~\protect\eref{eq.22} and~\protect\eref{eq.125c}. The
1\,$\sigma$ confidence region is also included for the latter.}
\label{fig:perform}
\end{figure}

The laboratory setup and the experimental details can be found in
reference~\cite{Marin06}. In this experiment, the optical window is not
part of a testing optical bench (OB), but is glued in a lateral side of
the \ltp OB Engineering Model, instead. This way, a double beam pass
across the window is forced: the laser light travels from the optical
bench through the OW to a \emph{dummy} mirror, faking a test mass; there,
it is reflected, sent back again across the OW and out to the OB. Such
setup proved to be compliant with the interferometer noise budget, showing
that the inclusion of the OW does not degrade the interferometer performance.
Two temperature sensors in the Titanium flange and one in the glass
were left in place to measure temperature values during long term runs.

No thermal disturbances were deliberately introduced, so the thermometers
only read environmental temperature fluctuations. We use
equation~\eref{eq.125c} to convert the temperature fluctuations spectral
density, $S_T^{1/2}(\omega)$, into a phasemeter spectral density,
$S_{\phi,T}^{1/2}(\omega)$. Thus,
\begin{equation}
 S_{\phi,T}^{1/2}(\omega) = 2\;
 |\widetilde{G}(\omega,\alpha,\beta,\delta)|\;S_T^{1/2}(\omega)
 \label{eq.22}
\end{equation}
where the numerical factor 2 is required to account for the double
passage of the laser beam through the window in this case. We assume
both passages are completely correlated, given the extremely small
time scale of their occurrence compared to thermal reaction times.
Spectral densities are therefore linearly added.

Results obtained in a typical run with that setup are plotted in
figure~\ref{fig:perform} using a MATLAB package being developed
\emph{ex professo} for the \ltp Data Analysis \cite{ltpda}.
The left panel shows temperature fluctuations measured in the Titanium
flange. As can be seen, these reach the front end electronics
(FEE) floor noise in the higher frequency region of the measuring
bandwidth, while keeping slightly above the \ltp maximum temperature
fluctuations requirements limit in the lower frequencies,
around~1~mHz~\cite{lastcqg}. This is in fact a worst case condition,
since the temperature power spectrum decreases as frequency increases,
and thus if the \ltp temperature requirement is reached at the lower
frequency range then the rest of the spectrum will naturally follow
a descending curve like the one shown in figure~\ref{fig:perform}.

The phasemeter fluctuations spectrum is however below the required noise
level, as we can see in the blue curve of the right panel. The temperature
fluctuations spectral data in the left panel are then submitted to the
algorithm, equation~\eref{eq.22}, and the result is the magenta curve
displayed in the right panel.

The low coupling to thermal disturbances implied by
$\widetilde{G}(\omega,\alpha,\beta,\delta)$ causes the thermal contribution
to only represent 5\,\% of the phasemeter noise at 1~mHz, and about
0.5\,\% of the \ltp goal. We thus feel reassured that there is still
considerable margin here.

\section{Continuous time models}
\label{sec.6}

On account of the empirical results reported in section~\ref{sec.3}, and of
the remarkable accuracy with which \emph{both} DLR and \textsl{ARMA} models
fit the experimental data \mbox{---notwithstanding} their completely different
nature---, we now try to shed some light into the kind of processes which
take place in the system.

For this, we attempt to picture the \textsl{ARMA}(2,1) model relating the
phase readout of the interferometer $\phi[n]$ and the temperature at the
Titanium flange $T_{\rm Ti}[n]$ as the digital implementation of some
\emph{analog} physical process. The starting point is of course the
digital algorithm, equation~\eref{eq.7}, which in this case is given by
\begin{equation}
 G(q,\alpha,\beta,\delta) = \alpha\,\frac{1-q^{-1}}{1+\beta q^{-1}} +
 \frac{\delta}{1+\beta q^{-1}}
\label{eq.23}
\end{equation}
where $q\/$ is the shift operator of equation~\eref{eq.8}. The recursive
form of the process thus defined is therefore
\begin{equation}
 \phi[n] + \beta\phi[n-1] = \alpha\left\{T_{\rm Ti}[n]-T_{\rm Ti}[n-1]\right\}
 + \delta\,T_{\rm Ti}[n]
 \label{eq.13}
\end{equation}
and can be regarded as the digital implementation of a first order
continuous time filter, governed by a first order differential equation:
\begin{equation}
 \dot{\phi}(t) + \tau^{-1}\,\phi(t) = A\,\dot{T}_{\rm Ti}(t)
 + B\,T_{\rm Ti}(t)\ ,\quad
 \left(\dot{}\equiv\frac{d}{dt}\right)
 \label{eq.14}
\end{equation}
where $\tau\/$ is the characteristic time constant of the analog filter, and
$A$ and $B\/$ are scale factors, respectively weighing the contributions of
the temperature's time variation rate and the temperature itself to the phase
shift effect. We have assumed the $T_{\rm Ti}(t)$ dependence in the rhs
of~\eref{eq.14} in line with the fit result expressed by the rhs
of~\eref{eq.13}.

If the time constant $\tau\/$ is much larger than the sampling time
$\Delta t$ implicit in equation~\eref{eq.13} then we can approximate
time derivatives by
\begin{equation}
 \dot{\phi}(t)\simeq\frac{\phi(t)-\phi(t-\Delta t)}{\Delta t}
 \label{eq.15}
\end{equation}
and, \emph{mutatis mutandi}, the same for $T_{\rm Ti}(t)$. Taking
$t\/$\,=\,$n\Delta t$ for the \mbox{timing of the $n\/$-th} sample,
and using the natural notation $\phi[n]$\,$\equiv$\,$\phi(n\Delta t)$,
equation~\eref{eq.14} is approximated~by
\begin{eqnarray}
 \hspace*{-3em}
 \phi[n]-\left(1+\frac{\Delta t}{\tau}\right)^{-1}\,\phi[n-1] & = &
 A\,\left(1+\frac{\Delta t}{\tau}\right)^{-1}\,\left\{
 T_{\rm Ti}[n]-T_{\rm Ti}[n-1]\right\} \nonumber \\
 & + & B\,\Delta t \left(1+\frac{\Delta t}{\tau}\right)^{-1}\, T_{\rm Ti}[n]
 \label{eq.16}
\end{eqnarray}

This can be readily compared to equation~\eref{eq.13} to obtain
\begin{equation}
 \hspace*{-4em}
 \beta = -\left(1+\frac{\Delta t}{\tau}\right)^{-1}\ ,\quad
 \alpha = A\,\left(1+\frac{\Delta t}{\tau}\right)^{-1},\quad
 \delta = B\,\Delta t\,\left(1+\frac{\Delta t}{\tau}\right)^{-1} 
\label{eq.17}
\end{equation}

$\beta$ is seen to have a value very close to $-1$ (Table~\ref{tab.1}),
or $\beta$\,=\,$-(1-\eta)$ with $\eta$\,$<$\,10$^{-2}$ comfortably in
all cases. Hence $\tau$\,$\simeq$\,$\Delta t/\eta$, i.e.,
$\tau$\,$\gg$\,$\Delta t$, which \emph{a posteriori} justifies the
approximation leading to equation~\eref{eq.16}.

The formal solution to equation~\eref{eq.14} can be easily written down.
After initial transients die out, the phase is given by
\begin{equation}
 \phi(t) = A\,T_{\rm Ti}(t) +
 (B-A)\,\tau^{-1}\int_0^t\,e^{-(t-t')/\tau}\,T_{\rm Ti}(t')\,dt'
 \label{eq.18}
\end{equation}

The meaning of this filter equation is better understood if we recast it
in frequency domain:
\begin{equation}
 \widetilde{\phi}(\omega) = \left[A\,\frac{i\omega\tau}{1+i\omega\tau}
 + B\,\frac{\tau}{1+i\omega\tau}\right]\,\widetilde{T}_{\rm Ti}(\omega)
 \label{eq.18c}
\end{equation}

This equation shows again that the analog process is also the superposition
of two contributions: a \emph{high-pass} filter proportional to $A$, and a
\emph{low-pass} contribution proportional to $B$: the first arises in
equation~\eref{eq.13} due to the Titanium temperature derivative, while
the second appears related to the term proportional to the Titanium
absolute temperature. This split dependence of the OW response to
temperature pulses points to two different physical thermal processes
affecting the glass, as already discussed in section~\ref{sec.4-1}.

We can now make use of equations~\eref{eq.17} to identify the coefficients
$A$ and $B\/$ in terms of the fit parameter values of Table~\ref{tab.1}.
Taking $\Delta t/\tau$\,$\ll$\,1, we find that $A\simeq\alpha$, and
$B\simeq\delta/\Delta t$. In addition, we can take advantage of the
relationship $\alpha\simeq p_1$ between the auto-regressive and the
DLR model parameters to obtain an expression relating both models.
Accordingly, equation~\eref{eq.18} can be rewritten as
\begin{equation}
 \phi(t)\simeq p_1\,T_{\rm Ti}(t) +
 (\delta/\Delta t-p_1)\,\tau^{-1}\,\int_0^t\,e^{-(t-t')/\tau}\,
 T_{\rm Ti}(t')\,dt'
 \label{eq.18b}
\end{equation}

If we go back to the DLR fit formula, equation~\eref{eq.2}, the following
expression ensues:
\begin{equation}
 T_{\rm Glass}(t) \simeq - \frac{p_1}{p_2}\,\tau^{-1}\,
 \int_0^t\,e^{-(t-t')/\tau}\,T_{\rm Ti}(t')\,dt'
 \label{eq.19}
\end{equation}
after the term $\delta/\Delta t\/$ has been been safely neglected in front
of $p_1$. We thus see that temperatures in the Titanium flange and in the
OW glass are related by a low-pass with a time constant, $\tau\/$, of a few
hundred seconds ---note that $p_1$ and $p_2$ have different signs,
Table~\ref{tab.1}.

It must be recalled that this relationship emerges out of the good quality
of the fits by both DLR and \textsl{ARMA}(2,1) models, and is key to
understanding why \emph{only} the Titanium gauge is required to make
a good prediction of the OW response to temperature variations, as will
be required in flight. The \emph{physical reason} for the observed
relationship between temperatures is to be sought in the properties
of the interface between the Titanium and the glass in the OW.

\section{Conclusions}
\label{sec.7}

While the optical window is a crucial element in the \ltp optical
Metrology system, it thankfully appears that it is quite stable to
temperature fluctuation noise ---so far as the latter is compliant
with mission environmental requirements. The present paper contains
a rather thorough analysis of such behaviour, based on experimental
data gathered through different runs of on-ground laboratory measurements.

Our main purpose was to prepare for thermal diagnostics analysis tools
in flight, and to gain as much understanding of the underlying physical
processes as possible. This means we need to know how noisy data retrieved
by thermometers can be converted into phasemeter fluctuations, thereby
quantifying the contribution of temperature random variations to the
total mission noise budget ---which is the ultimate objective of \lpf
in preparation for \lisa. Our most relevant finding is the discovery
that temperature readings in the Titanium flange embracing the OW
plane-parallel plate relate to phase values through an \textsl{ARMA}(2,1)
transfer function. Although this is the result of numerical analysis, hence
lends itself to parameter estimation variances, it appears to be considerably
robust.

The analysis has shown that the \textsl{ARMA}(2,1) process naturally splits
up as the sum of a high-pass and a low-pass process, each of them with
significantly different relative weights which result in the high-pass
dominating above $\sim$1~mHz, while the low-pass takes over in the lower
\lisa frequency band, i.e., at 0.1~mHz and below. A major achievement
of the analysis has been the identification of the physical processes
responsible for this behaviour: mechanical stresses ---induced by
differential thermal expansion of metal and glass--- are associated
to the high pass term, while $d\phi/dT|_{\rm free}$ effects account
for the low-pass.

We consider the analysis presented here as rather complete in some of its
essential traits. But there are still open issues which call for further
study. For example, heater generation of test signals must be monitored by
temperature sensors close to the activated heaters ---due to lag effects
in remoter spots--- for the procedures described herein to be fully
operative. This raises some caveats regarding full applicability of
the noise projection algorithms, as the sources of heat dominating a
given temperature reading may not be clear in \ltp science operation
mode.

A more global tool for full \ltp thermal diagnostics, which takes into
account the specific features of each individual part of the system must
be assembled. Research on this is currently underway which will reported
in due course.

\ack

Support for this work came from Project ESP2004-01647 of Plan
Nacional del Espacio of the Spanish Ministry of Education and Science
(MEC). MN acknowledges a grant from Generalitat de Catalunya, and JS
a grant from MEC.

\end{document}